# Atomic Layer Deposition of Cerium Dioxide Film on TiN and Si Substrates: Structural and Chemical Properties

Silvia Vangelista[1], Rossella Piagge[2], Satu Ek[3], Tiina Sarnet[3], Gabriella Ghidini[2] and Alessio Lamperti[1]

[1] CNR-IMM - MDM Laboratory, Via C. Olivetti 2, Agrate Brianza (MB) I-20864 Italy
[2] STMicroelectronics, Via C. Olivetti 2, Agrate Brianza (MB) I-20864 Italy
[3] Picosun Oy, Tietotie 3, Espoo FI-02150 Finland

**ABSTRACT**

Cerium dioxide ($CeO_2$) thin films were deposited by atomic layer deposition (ALD) on both Si and TiN substrates. The ALD growth produces $CeO_2$ cubic polycrystalline films on both substrates. However, the films show a preferential orientation along <200> crystallographic direction for $CeO_2$/Si or <111> for $CeO_2$/TiN. In correspondence, we measure a relative concentration of $Ce^{3+}$ equals to 22.0% in $CeO_2$/Si and around 18% in $CeO_2$/TiN, by X-ray photoelectron spectroscopy. Such values indicate the presence of oxygen vacancies in the films. Our results extend the knowledge on the structural and chemical properties of ALD-deposited $CeO_2$ either on Si or TiN substrates, underlying films differences and similarities, thus contributing to boost the use of $CeO_2$ through ALD deposition as foreseen in a wide number of applications.

**INTRODUCTION**

In the last years, $CeO_2$ based materials have attracted much attention due to their wide use in many application areas such as catalysis, hydrogen production, gas sensing and electrodes in fuel cells [1,2,3]. In microelectronics, $CeO_2$ has been considered a high κ-gate oxide candidate since its moderate band gap (3–3.6 eV), high dielectric constant (κ: 23–26) and high dielectric strength (~2.6 MV cm$^{-1}$) [4]. $CeO_2$ is also suitable for Si MOS devices due to its small lattice mismatch with Si and low interface-state density (~ $10^{11}$ cm$^{-2}$eV$^{-1}$) [5]. For catalytic applications $CeO_2$ has to be grown on metallic substrates, but few studies exist on cerium oxide film deposited on metals and most of them have the aim to depict the epitaxial growth [6 and references therein]. $CeO_2$ deposition is achieved by using a variety of growth techniques [7, 8, 9,10]. When highly stoichiometric oxide is required, atomic layer deposition (ALD) is the most suitable technique. ALD is a self-limiting technique where thin films deposition can be performed at lower temperatures than other vacuum deposition techniques such as pulsed laser deposition or CVD, and the process condition guarantee low thermal budget, no or limited interdiffusion phenomena, and the possibility to use temperature-sensitive substrates. ALD deposition of $CeO_2$ on Si or Si/$SiO_2$ substrates has been explored by using different precursors [11,12,13,14], obtaining as-deposited films with polycrystalline structure [12,13], which ultimately influences the dielectric behavior [14]. However, ALD of $CeO_2$ on metal substrates has not been explored yet. In this work, we the aim to a deep understanding of the structural properties of ALD-deposited $CeO_2$ either on Si or TiN substrate, underlying differences and similarities between the obtained film properties. Our results will contribute to give the basis for the use of $CeO_2$ through ALD deposition as foreseen in a wide number of applications, such as in microelectronics and in catalysis.

**EXPERIMENT**

Cerium dioxide thin films were deposited in a Picosun R-200 Advanced ALD reactor by Picosun Oy. The precursor $Ce(thd)_4$ (thd = 2,2,6,6-tetramethyl-3,5-heptanedione) was evaporated at 140 °C and ozone has been used as oxidizing agent (concentration 19%). The deposition temperature was 250 °C, the pulse times for $Ce(thd)_4$ and ozone were 1.0 s and 2.5 s respectively, and the purge time with nitrogen after pulses was 1.5 s. The targeted oxide thickness was 25 nm. The deposition was done on both Si and TiN-coated 7×7 $cm^2$ Si substrates simultaneously. X-ray diffraction at grazing incidence (GIXRD) have been performed using an XRD3000 diffractometer (Italstructure) with monochromated X-ray Cu Kα radiation (wavelength 0.154 nm) and beam size of 0.1×6 mm. MAUD program has been used for Rietveld refinement of GIXRD patterns [15]. The compositional depth profile of the $CeO_2$/TiN/Si and $CeO_2$/Si structures was investigated by Time of Flight-Secondary Ion Mass Spectrometry (ToF-SIMS) using $Cs^+$ ion beam (energy of 0.5 keV, ion current 38.5 nA) for sputtering a 200 μm × 200 μm area, and $Ga^+$ ion beam (25 keV, 2.7 pA) for analysis over a 50 μm × 50 μm area centred on the sputtered crater and therein collecting secondary negative ions. X-ray photoelectron spectroscopy (XPS) measurements were performed on a PHI 5600 instrument (Physics Electronics Inc.) with a monochromatic AlKα X-ray source (1486.6 eV) and a concentric hemispherical analyser. The spectra were collected at a take-off angle of 45° and band-pass energy of 23.5 eV. The instrument resolution is 0.5 eV. The experimental data were fitted with Voigt peaks and Shirley background by using XPSPeak4.1 program [16].

**DISCUSSION**

The values of thickness, roughness and electronic density have been checked by X-Ray Reflectivity (XRR). $CeO_2$ thickness is close to the target (25 nm) in both cases. The surface roughness is limited to about 1 nm. On the top of the as-deposited films the absorption of humidity is detected, a well-known phenomenon for $CeO_2$ [17], due to its intrinsic hygroscopicity. The interface roughness is higher in the $CeO_2$/TiN sample than in the $CeO_2$/Si sample and, in addition, a very thin oxidation of TiN surface is detected. The crystallinity of the films was examined by GIXRD at fixed incidence angle of 2°. Fig. 1 shows the GIXRD patterns of as-deposited $CeO_2$ film on Si and TiN substrates. Both films are polycrystalline, exhibiting the typical peaks of a face centered cubic (fcc) structure, as reported in the crystallographic database [18]. No signal coming from the TiN layer can be detected (TiN thickness is 7.2 nm from XRR). In particular, we found that the relative intensity of $CeO_2$ main peaks (*(111)* and *(200)* peaks) is useful to indicate some preferential orientation of the $CeO_2$ films. This becomes even more clear by performing Rietveld refinement of the GIXRD data (dots on top of the experimental data - Fig. 1) assuming the powder pattern ICSD 621705 [18] as initial input.

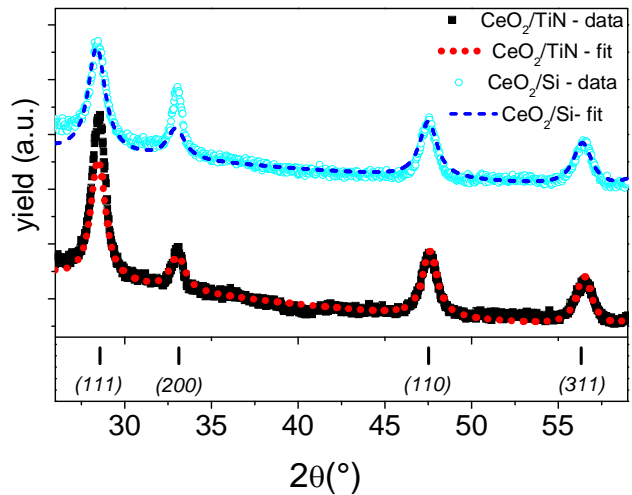

**Figure 1.** GIXRD experimental data and fitting of as-deposited $CeO_2$ on TiN and on Si substrates. In the lower panel the peaks positions are reported, taken from diffraction powder pattern ICSD no. 621705.

For $CeO_2$ deposited on TiN (lower black squares) the $I_{111}/I_{200}$ intensity ratio is higher than that of the powder spectrum, while for the $CeO_2$ deposited on Si (upper cyan circles) the $I_{111}/I_{200}$ intensity ratio is lower than that of the powder diffraction pattern, the latter observation in accordance with ref. [19]. These findings suggest that the $CeO_2$ growth is influenced by the different nature of the surface/layer onto which it is deposited. It is likely that the presence of a higher roughness at the $CeO_2$/TiN interface with respect to the case of $CeO_2/SiO_2$ could contribute in a change to the energy requirement for the nucleation of crystals with *(111)* orientation. It has to be underlined that the atomistic simulations [20, 21] show that the formation of crystallites exposing *(100)* surfaces is unstable because of the dipole generated perpendicular to it. However, the formation of polycrystalline $CeO_2$ with all crystalline orientations is well-demonstrated experimentally by using different growth methods [12, 13], and more generally it can be justified considering defects or charge compensating species, i.e. oxygen vacancies due to $Ce^{3+}$: $Ce^{4+}$ substitution. In our case, the contaminants C and OH from the precursor can act as the source of defects, or there could be oxygen vacancies due to a different reactivity of TiN versus Si.

In order to understand the presence and the nature of the contaminants in the two films we performed ToF-SIMS chemical depth analysis (Fig. 2(a) and 2(b)). First, we can observe that the $CeO_2$/Si interface is very sharp (Fig. 2(a)), with the existence of a thin silicon native oxide layer in between. On the other side, the $CeO_2$/TiN interface is slight broad (Fig. 2(b)), probably due to the roughness of the $CeO_2$/TiN interface rather than representing a sign of diffusion of Ti into the $CeO_2$ film, also considering the low temperature of the growth. Moreover, ToF-SIMS allows to report the signals coming from C and OH, which are the typical spurious elements incorporated during an ALD deposition when Me(thd) and $O_3$ precursors are used for the growth [12]. By comparing the levels of the signals between the two samples we observe that, while OH shows almost the same intensity, C content (only few atomic percent) is comparatively lower in $CeO_2$/TiN than in $CeO_2$/Si. In any case, it is important to underline the presence of carbon, eventually in presence also of shortage of oxygen, since it can induce the reduction of $Ce^{4+}$ into $Ce^{3+}$, which is fundamental in the use of ceria as catalytic converter. The presence of $Ce^{3+}$ can be related to the presence of $Ce_2O_3$-like phase, or to a highly non-stoichiometric $CeO_{2-x}$ phase [22]. We lean towards this second hypothesis, since we do not have any other indication of $Ce_2O_3$: indeed, we cannot detect $Ce_2O_3$ diffraction peaks in the XRD diffraction patterns.

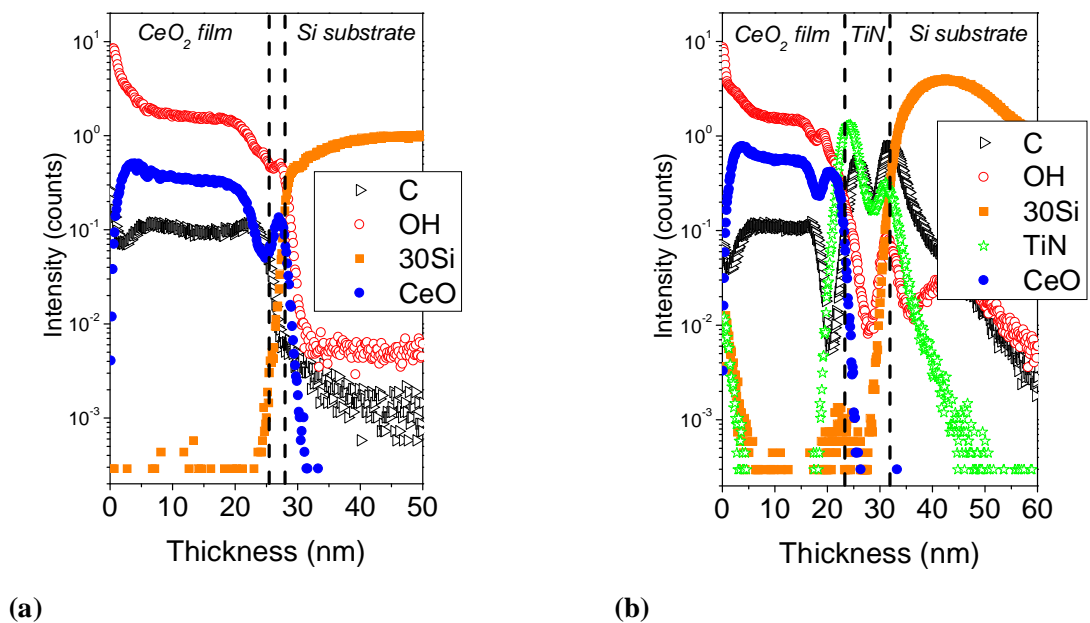

(a)                                     (b)

**Figure 2.** (a) ToF-SIMS profile of $CeO_2$ film deposited on Si and (b) on TiN.

The off-stoichiometry of the oxide can be estimated from the XPS analysis. Despite the technique probing volume is limited to the outermost part of the $CeO_2$ layers, which can convey of the interaction with the atmosphere, it can supply an indication of the chemical stoichiometry of the whole oxide layer. Fig. 3(a) and 3(b) shows the *Ce3d* and *O1s* spectra respectively of both films ($CeO_2$ on Si –upper panel, and on TiN – lower panel). The figures show also the fit obtained from the deconvolution of the spectra into the different components, similarly to that reported in refs. [19, 22]. The analysis of the *Ce3d* spectrum is particularly complex since the deconvolution had to consider 5 different doublets: three related to the $Ce^{4+}$ doublets and the other two related to the $Ce^{3+}$ doublets. By comparing the *Ce3d* (and *O1s*) spectra of the two films, it is evident a shift of one signal respect to the other, revealing the different chemical environment in the two samples (the spectra have been aligned to *C1s* peak at 284.6 eV). From the different components area of the *Ce3d* spectrum we estimated the $Ce^{3+}$ and $Ce^{4+}$ relative concentration: in $CeO_2$/Si sample is 21.8%, while in $CeO_2$/TiN sample is 18.0%. Similarly, considering *O1s* region, 3 components are identified and assigned to oxygen bonded to $Ce^{4+}$,

$Ce^{3+}$ and $CO_x:H_2O$ complexes from the environment [23]. From the *O1s* region, the $Ce^{3+}$ and $Ce^{4+}$ relative concentration is 21.0% $Ce^{3+}$ in $CeO_2/Si$ and in 18.7% $Ce^{3+}$ in $CeO_2/TiN$, in agreement with those calculated from *Ce3d*.

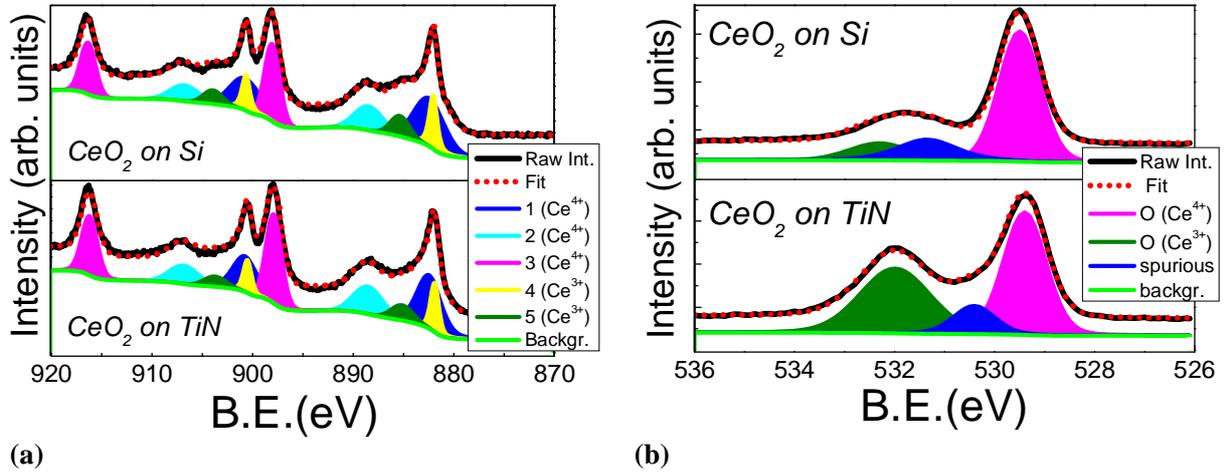

**Figure 3.** (a) *Ce3d* XPS spectra collected from $CeO_2$ on Si (upper panel) and on TiN (lower panel). The fit components of the $Ce^{3+}$ and $Ce^{4+}$ related doublets are shown; (b) *O1s* XPS spectra from $CeO_2$ on Si (upper panel) and on TiN (lower panel). The 3 fit components are shown.

It is well-known that $Ce^{3+}$ oxidation state is preferentially located at the surface of the oxide film [22] due to the interaction with organics in the air, thus we can presume that the $Ce^{3+}$ percentage represents the upper limit of the off-stoichiometry for $CeO_2$ layers in both samples. Considering the calculated concentrations for $Ce^{3+}$ and $Ce^{4+}$, we obtain an indication of ~10% off-stoichiometry, which may suggest that $Ce^{3+}$ in the film is possibly related with the presence of oxygen vacancies, while $Ce_2O_3$ amorphous phase cannot be completely excluded. Nonetheless, the difference of $Ce^{3+}$ amount between $CeO_2/Si$ and $CeO_2/TiN$, despite small, could be sufficient for inducing a different preferential crystallization in ceria films, as observed by XRD.

**CONCLUSIONS**

In this work we report about a structural and chemical characterization of $CeO_2$ thin films deposited by ALD on both Si and TiN substrates. This oxide has a cubic polycrystalline structure, with a preferred orientation of the grains depending on the substrate. Indeed, we observe an enhancement of the diffraction signal connected to the *(100)* orientation of the grains in the case of Si substrate, compared to the enhancement of the diffraction signal related to the *(111)* direction for the TiN substrate. The not-favoured exposure of *(100)* planes for strongly ionic oxides like $CeO_2$ can be explained considering the higher interface $CeO_2/TiN$ roughness compared to $CeO_2/Si$ one, which can contribute in a change to the energy requirement for the nucleation of crystals with different orientation. Further, from XPS we deduced a slight different $Ce^{3+}$ concentration in the two samples, sufficient to induce the observed different structural features, since the presence of defects or charge compensating species, i.e. oxygen vacancy or

$Ce^{3+}$, can change the energy requirement. Our study evidences how the substrate plays a role in the growth of $CeO_2$ films by ALD, contributing to boost the use of ALD based processes targeting a wide range of applications and a viable route for ceria oxide engineering to address optimal properties for specific applications, such as in microelectronics and catalysis.

## ACKNOWLEDGMENTS

The authors would thank Dr. S. Spiga for fruitful discussions on the manuscript content, D. Brazzelli, F. Toia and D. Piccolo (ST-Agrate) for discussions on sample preparation. Partial financial support by ECSEL-JU R2POWER300 project under grant agreement n.653933 is also acknowledged.